\documentclass[conference]{IEEEtran}
\IEEEoverridecommandlockouts
\usepackage[ruled,vlined]{algorithm2e}
\usepackage[table]{xcolor}

\newcommand{\xvar}[1]{\textit{#1}}

\newcommand{\xvbox}[2]{\makebox[#1][l]{#2}}

\SetAlFnt{\footnotesize}

\SetCommentSty{xCommentSty}

\SetNoFillComment

\SetNlSty{mynlfont}{}{} 

\LinesNumbered

\SetSideCommentRight

\DontPrintSemicolon

\RestyleAlgo{algoruled}

\SetKwProg{Fn}{def}{}{}{}

\SetKwInOut{Input}{In}
\SetKwInOut{Output}{Out}

\makeatletter

\newcommand{\checknextarg}{\@ifnextchar\bgroup{\gobblenextarg}{}}
\newcommand{\gobblenextarg}[1]{ $\leftarrow$ $#1$\@ifnextchar\bgroup{\gobblenextarg}{}}

\usepackage{amsmath,amssymb,mathrsfs,mathtools,lipsum,amsthm}
\theoremstyle{definition}
\newtheorem{theorem}{Theorem}[section]

\newtheorem{lemma}[theorem]{Lemma}
\newtheorem{definition}{Definition}[section]

\newcommand*{\field}[1]{\mathbb{#1}}%
\DeclareMathOperator*{\argmax}{arg\,max}

\DeclareMathOperator*{\subjectto}{s.t.}
\DeclarePairedDelimiter{\ceil}{\lceil}{\rceil}
\usepackage{cuted}
\usepackage{color}
\usepackage{setspace}

\newcommand{\ie}{{\it i.e.}}
\newcommand{\eg}{{\it e.g.}}
\newcommand{\etc}{{\it etc}}

\newcommand{\specialcell}[2][c]{%
	\begin{tabular}[#1]{@{}c@{}}#2\end{tabular}}

\newcounter{magicrownumbers}

\newcommand{\RNum}[1]{\uppercase\expandafter{\romannumeral #1\relax}}
\newcommand{\Srel}[1]{\mathrel{{}^{#1}\kern-\scriptspace}}

\usepackage{array}
\newcolumntype{L}[1]{>{\raggedright\let\newline\\\arraybackslash\hspace{0pt}}m{#1}}
\newcolumntype{C}[1]{>{\centering\let\newline\\\arraybackslash\hspace{0pt}}m{#1}}
\newcolumntype{R}[1]{>{\raggedleft\let\newline\\\arraybackslash\hspace{0pt}}m{#1}}

\usepackage{enumitem}
\usepackage{textcomp}
\usepackage{booktabs}

\usepackage{graphicx}
\graphicspath{ {images/} }

\usepackage{caption}
\usepackage{subcaption}

\usepackage{wrapfig}
\newcommand{\tuple}[1]{\ensuremath{\left \langle #1 \right \rangle }}

\usepackage{upgreek}
\usepackage{bm} 

\usepackage{cuted}
\usepackage{multirow}
\usepackage{listings}
\definecolor{dkgreen}{rgb}{0,0.6,0}
\definecolor{gray}{rgb}{0.5,0.5,0.5}
\definecolor{mauve}{rgb}{0.58,0,0.82}
\usepackage{fontenc}
\usepackage{inputenc}

\begin{document}

\title{\LARGE SODA: A Semantics-Aware Optimization Framework for Data-Intensive Applications Using Hybrid Program Analysis}



\author{
    \IEEEauthorblockN{Bingbing Rao\IEEEauthorrefmark{1}, Zixia Liu\IEEEauthorrefmark{1}, Hong Zhang\IEEEauthorrefmark{2}, Siyang Lu\IEEEauthorrefmark{3}, Liqiang Wang\IEEEauthorrefmark{1}}
    \IEEEauthorblockA{\IEEEauthorrefmark{1}{\it Department of Computer Science}, University of Central Florida, Orlando, FL, USA}
    \IEEEauthorblockA{\IEEEauthorrefmark{2}{\it School of Cyber Security and Computer}, Hebei University, Baoding, Heibei, China\\}
    \IEEEauthorblockA{\IEEEauthorrefmark{3}{\it School of Computer and Information Technology}, Beijing Jiaotong University, Beijing, China}
}




\maketitle

\thispagestyle{plain}
\pagestyle{plain}

\begin{abstract}
In the era of data explosion, a growing number of data-intensive computing frameworks, such as Apache Hadoop and Spark, have been proposed to handle the massive volume of unstructured data in parallel. Since programming models provided by these frameworks allow users to specify complex and diversified user-defined functions (UDFs) with predefined operations, the grand challenge of tuning up entire system performance arises if programmers do not fully understand the semantics of code, data, and runtime systems. In this paper, we design a holistic {\it \underline{\bf s}emantics-aware \underline{\bf o}ptimization for \underline{\bf d}ata-intensive applications using hybrid program \underline{\bf a}nalysis} (SODA) to assist programmers to tune performance issues. SODA is a two-phase framework: the {\it offline} phase is a {\it static analysis} that analyzes code and performance profiling data from the online phase of prior executions to generate a parameterized and instrumented application; the {\it online} phase is a {\it dynamic analysis} that keeps track of the application's execution and collects runtime information of data and system. Extensive experimental results on four real-world Spark applications show that SODA can gain up to 60\%, 10\%, 8\%, faster than its original implementation, with the three proposed optimization strategies, \ie, cache management, operation reordering, and element pruning, respectively. 
\end{abstract}


\section{Introduction}
\label{sec:intro} 


With data explosion in many domains, such as Internet of Things (IoT)~\cite{ashton2009internet}, scientific experiments~\cite{demchenko2013addressing, subramanian2011rapid, zhang2015dart}, e-commerce, and social media~\cite{chung2016dynamic, chung2019interaction}, people are facing an increasing number of obstacles concerning data processing and analytics on the sheer size of these unstructured data. These obstacles include interactive computing 
and user-specific element-wise data transformations~\cite{alexandrov2019representations}. To break through these dilemmas, a growing number of data-intensive computing frameworks have been proposed, such as MapReduce~\cite{dean2008mapreduce}, Apache Hadoop~\cite{hadoop2009hadoop}, and Spark~\cite{zaharia2010spark}. Generally, a mainstream approach to gain computing capability and scalability behind these platforms is to distribute data and computations across a cluster of nodes so that a large volume of data can be processed in a parallel and robust manner within a reasonable time~\cite{rao2017survey, zomaya2017handbook}. The successes of these frameworks owe to their MapReduce-like programming models, which are further based on data distribution techniques~(\eg, Resilient Distributed Dataset (RDD) in Apache Spark~\cite{zaharia2012resilient}), and high-order functions~(\eg, $map$, $reduce$, $filter$) that can take user-defined functions (UDFs) as arguments. The semantics of these high-order functions facilitate data-parallelism to manipulate datasets in an element-wise way while UDFs are applied to each element to produce the desired result. 

Despite these advantages, an endeavor to improve the performance of data-intensive applications exhibits a few challenging issues. 1) Usually, unstructured data expose less information about their schema if without metadata or annotation provided by programmers or help from runtime profiling tools. 2) It is difficult to apply conventional database-style optimizations on unstructured data directly, such as relational algebraic reordering and filter pushdown, since the programming models of current data-intensive computing platforms lack information about data schema~\cite{alexandrov2019representations}. 3) Although Spark can process raw unstructured data directly using DataFrame or Dataset APIs, it needs to parse them before performing transformations (e.g, Map) and actions (ReduceByKey). Particularly, these applications can spend 80-90\% of the entire executing time in data parsing~\cite{palkar2018filter}. 4) Programming models usually treat UDFs as black-boxes and their semantics are therefore hidden from the system, resulting in insufficient information for further optimization~\cite{guo2012spotting, hueske2012opening, rheinlander2015sofa, rheinlander2017optimization}. 5) Runtime factors are not fully utilized to tune the performance of a specific operation's execution, such as cache management~\cite{perez2018reference, yu2017lrc}. Therefore, it is vital to integrate program semantics, data property and runtime factors to improve the performance of data-intensive applications since pure static optimizations are either limited or impossible if without efficient profiling information about data and runtime systems.

In this paper, we present a two-phase semantics-aware approach to optimize data-intensive applications combining static and dynamic program analysis. The first phase is an {\it offline static analysis}. Source code and performance log collected in prior executions are analyzed to refactor code by applying three kinds of optimizations: cache management, operation reordering and element pruning. The offline phase is developed as a compiler plugin of the host development languages (\eg, Scala, Java). However, not all performance issues can be fixed in the offline phase, while some may need support from system runtime information, such as intermediate data size, memory usage and execution time of operations. The second phase is an {\it online dynamic analysis} to obtain the required runtime information, where we instantiate a parameterized framework based on the instrumentation generated in the offline phase to trace applications' execution and extract profiling information concerning data and system status. Unless otherwise specified, we shall limit our discussion to the context of Apache Spark and demonstrate its effectiveness on Spark applications in the following sections. Nevertheless, the proposed approach is general and can be extended to other data-intensive platforms. To the best of our knowledge, our implementation is the first compiler plugin to help users optimize data-intensive applications. The major contributions of this work are summarized as follows:
\begin{itemize}
    \item A framework on Spark using hybrid program analysis is proposed to optimize the performance of data-intensive applications on unstructured data at the code level.
    
    \item We design approaches to detect three kinds of performance issues from the perspectives of data, code and system, respectively: cache management, operation reordering and element pruning.
    
    \item SODA prototypes maximize expected caching gain offline by reducing it to a convex-concave relaxation problem and leverages Pipage Rounding approximation algorithm to construct a probabilistic cache policy within $1 - 1/e$ factor from the optimal, in expectation. 
    
    \item A novel reference measurement, named Global Execution Distance, is proposed based on application workflow to narrow down search space in the Pipage Rounding algorithm. 
    
	\item A piggyback profiling tool is integrated with Spark internal metrics and event system to gather statistics for applications during runtime.
\end{itemize}

The rest of this paper is organized as follows. Section \ref{sec:overview} reviews performance problems and introduces the whole life cycle of SODA. A comprehensive discussion about the semantics-aware data model is provided in Section~\ref{sec:sem_model}. Section~\ref{sec:opt} discusses the philosophy behind three kinds of optimization strategies. The evaluation and experiments of SODA are illustrated in Section~\ref{sec:exper}. Section~\ref{sec:related_work} sketches related work. Finally, conclusions and future work are presented in Section~\ref{sec:conclusion}. 

\section{System Overview}
\label{sec:overview}

SODA is proposed as a two-phase framework, \ie, offline static analysis and online dynamic analysis, to interactively and semi-automatically assist programmers to scrutinize performance problems camouflaged in source code.

\begin{figure*}[ht!]
	\centering
	\includegraphics[scale=0.6]{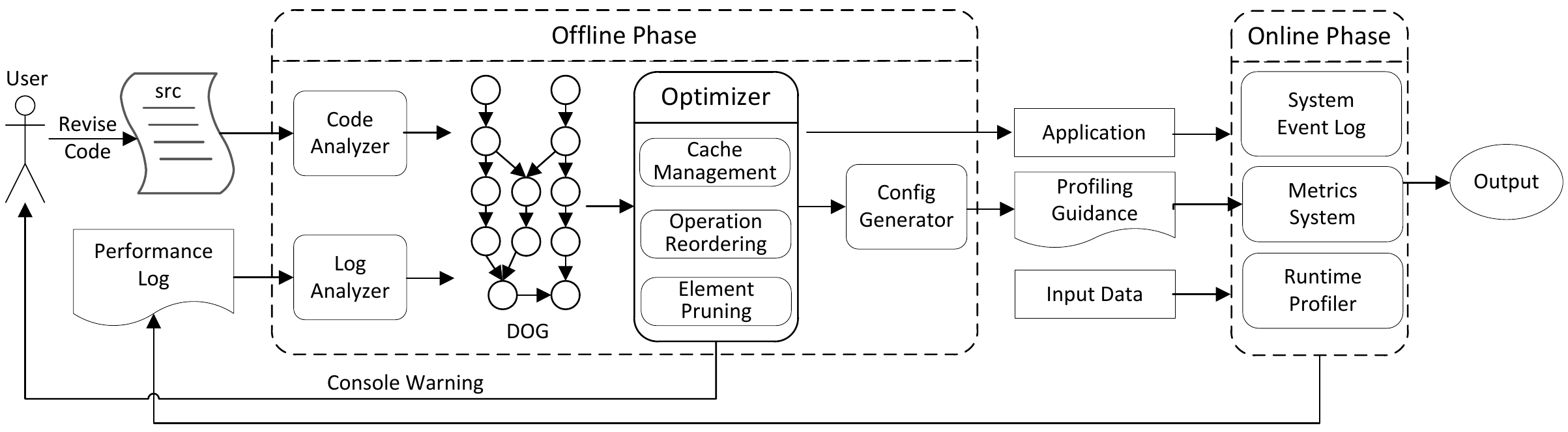}
	\caption{The full life cycle of Semantics-Aware Optimization Approach for Data-Intensive Applications (SODA).}
	\vspace{-6mm}
	\label{fig:arch}
\end{figure*}

\subsection{Performance Problems}
\label{sec:perf_prorblems}

SODA looks for three kinds of performance problems: Cache Management~(CM), Operation Reordering~(OR) and Element Pruning~(EP).

{\bf Cache Management (CM):} It is crucial to manage cache resource for these data analytics frameworks~\cite{saha2015apache, shinnar2012m3r, toshniwal2014storm, zaharia2010spark}, which leverage in-memory computing to speed up performance and bypass the hindrance of disk and network I/O. Within these systems, intermediate computing data block would be put in memory by default. There is a block management component to manage these blocks and determine when and which one is evicted from memory. Recently, a rich line of research work propose different data block reference measurements to improve cache hit, such as least recently used~(LRU), least reference count~(LRC)~\cite{yu2017lrc} and most reference distance~(MRD)~\cite{perez2018reference}. However, there remain two important factors that previous works have not taken into account, especially in Spark. 
\begin{enumerate}
    \item The executing order of all stages. This could impact system performance, especially for cache behaviors.
    \item Data block size. Data blocks with the same reference in LRU or other fancy measurements, might not all fit the memory at the same time, and therefore it raises the concern about cache priority regards to system performance.
\end{enumerate}

In addition, programmers may brutally persist the desired dataset in memory by invoking corresponding APIs (\ie, using the {\tt persist} (or {\tt cache}) method in Spark), resulting in a more complicated research problem. Therefore, an intelligent cache management mechanism using hybrid program analysis is needed to manage memory for efficiency.  In this paper, we propose a stage-level cache allocation strategy in a data-intensive system by reducing it to a convex optimization problem~\cite{boyd2004convex, ioannidis2016adaptive, yang2018intermediate}.

{\bf Operation Reordering(OR):} A data-intensive system usually supports a rich line of operations, such as {\tt map, reduce, filter, reduceByKey}, and {\tt join}. A developer may face a variety of executing plans assembled by a sequence of operations associated with UDFs to accomplish an application. Nonetheless, not all of these arrangements will yield identical performance. Therefore, it is crucial to orchestrate the operations in an appropriate order to bypass common pitfalls affecting performance significantly. For example, filter pushdown and join reorder are two common optimization strategies to improve the performance of relational algebra-based systems when handling structured data. As to process unstructured datasets, however, it is difficult to apply these conventional database-style techniques to systems using non-relational algebraic programming models. In this paper, we propose SODA to break through such kinds of dilemmas and extend these two strategies into a more general approach for processing unstructured data.

{\bf Element Pruning (EP):} It is common that not all portions of a dataset are used to produce output, which leads to a series of redundant I/O operations, such as Disk I/O for reading/writing data and network I/O for transferring data among computing nodes. These redundant operations can become more severe when processing unstructured data. Due to the lack of predefined schema of a given dataset, it is difficult to detect data workflow in a fine-grained granularity~(\ie, on attribute level), hence fails to identify unused data attributes. In particular, the redundant portion of the dataset may be a dominant barrier for performance when shuffling a huge size of data across networks in a data-intensive computing system.



\subsection{Architecture \& Background}

The framework of SODA includes offline and online phases, as shown in Figure~\ref{fig:arch}. The offline~(static) phase is developed as a compiler plugin of host programming languages~(\ie, Scala, Java), and analyzes source code~(src) and performance log about data and runtime factors to generate a nearly-optimized and parameterized program. Firstly, {\it Code Analyzer} analyzes source code with the help of a local compiler to construct a directed data operation graph (DOG), which represents the skeleton of an application. This graph comprises a set of nodes and edges, which denote operations and dataflows, respectively. In addition to static properties associated with corresponding operations, a group of dynamic profiling data is extracted by {\it Log Analyzer} from the performance log, which is accumulated in prior executions, including execution time, memory usage, input, and output data size of operations, runtime system status \etc. These information can be extracted from system log~\cite{lu2017log, lu2018detecting, lu2019ladra} and provided by our profiling tool using Javassist (Java Programming Assistant), a high-level bytecode instrumentation tool to instrument APIs of Spark to expose information needed~\cite{chiba1998javassist}. Next, three optimization strategies, \ie \; cache management, operation reordering and element pruning, are applied to assist users to scrutinize performance problems. When a problem is found, users would get informed about performance bugs from SODA and then refactor code. However, not all problems can be determined statically, it may need more information coming from executions. For example, SODA makes use of execution time and output size of operations to verify performance behavior and then create a global cache allocation strategy. To reduce system overhead resulting from the profiling process, {\it Config Generator} produces {\it Profiling Guidance}  to inform the online phase about which operations and what kinds of computational resources needing to be monitored. In the online phase, SODA initializes an application with a parameterized configuration based on {\it Profiling Guidance} and starts a piggyback listener residing in each worker and master node to collect runtime information about memory usage, data property and system configuration. The profiling data would be accumulated and then delivered back as a performance log to the offline phase for further optimizations. 

In this paper, SODA is implemented on Apache Spark and several real-world Spark applications are used as benchmarks to evaluate its effectiveness. Apache Spark~\cite{zaharia2012resilient} is an efficient and general engine for large-scale data processing that supports the scalability of MapReduce~\cite{dean2008mapreduce}. Its main abstraction, named Resilient Distributed Dataset (RDD)~\cite{zaharia2012resilient}, is a fault-tolerant and immutable collection of objects, which are logically partitioned across a cluster of computing nodes so that they can be manipulated in parallel. Spark's programming model provides two types of operations, {\it transformation} and {\it action}. A {\it transformation} creates a new RDD dataset from an existing one while an {\it action} returns a value to the driver program. The lazy feature of transformations enables Spark to run more efficiently since they do not compute their results immediately until action is invoked. An RDD has to be recomputed when invoking an action on it unless it is persisted in memory using the {\it persist (or cache)} method, which facilitates much faster access. Apache Spark automatically monitors cache usage on each node and drops out old data partitions in an LRU fashion by default. 
In the Spark execution model, a Spark application is divided into a group of jobs executed in a sequential order\footnote{Multiple jobs can run simultaneously if they were submitted from separate threads}, where a job is a parallel computation in response to a Spark action (\eg,\;{\it save}, {\it collect}); 
Within a {\it job}, multiple {\it stages} are generated and bounded by shuffle behaviors~(\eg, reduce), then runs in parallel if there is no data dependency among them; otherwise, they are scheduled sequentially. Internally, a stage is a physical execution unit consisting of several operations. The unit is further divided into tasks, which share identical code but run on different data partitions in parallel. Given that, we need a fine-grained profiling tool to analyze semantic properties in code as well as runtime factors, such as the evolution of data, the execution time of operations and system status, to narrow down the gap between the programming model and execution model).

\section{Semantics-Aware Data Model}
\label{sec:sem_model} 

We propose {\it semantics-aware data model} to keep track of the evolution of dataset(s). 


\subsection{Attribute-Based Data Abstraction}
\label{subsec:rec_data}

SODA parses and represents an unstructured dataset as a multiset of elements in which repetitive ones may be included, termed as $X = \{x_{1},...,x_{n}\}$, where $n$ is the number of elements. To exploit datasets deeper and provide more information to optimizations, SODA treats an element $x \in X$ as an ordered $m$-tuple: $x = \tuple{x[a_1],\ldots,x[{a_m}]}$, where $x[a_i]$ is the value(s) of an attribute $a_i$. One or two datasets can be manipulated by an operation (including user-defined function (UDF)) to generate a new dataset, where the operation can access and transform attributes of an element. Let $Y = X.op(f)$ denote that a new dataset $Y$ is generated by applying a unary operation $op$ (\eg, $map$, $reduce$, and $filter$) and its corresponding UDF $f$ to an input dataset $X$. Similarly, we can define binary operations. In the following discussion, we use unary operations to demonstrate our approach for simplicity, and the same idea can be applied to binary operations. 

To process such a transformation in static code analysis, SODA first models attributes of $X$ and $Y$ by analyzing their type information, as well as the input and output of $f$. Let $\beta(X)$, $\beta(Y)$ denote all extracted attributes of $X$ and $Y$, respectively. Next, SODA analyzes the source code of $f$ to create dataflows between $\beta(X)$ and $\beta(Y)$ at the level of the attribute.

\subsection{Primitive Operations}
\label{subsec:operation}

SODA defines six primitive operations to imitate common behaviors of a general data-intensive system, as shown in Table~\ref{tab:op_def}.

\begin{table*}[ht!]
	\centering
	\small
	\begin{tabular}{|c|l|l|}
		\hline
		Operation & Notation & Examples in Apache Spark\\\hline
		{\it Map} &  $Map: X \times f \mapsto Z$  &{\tt map, flatmap, mapValues,mapPartions} \\ \hline
		{\it Filter}  &  $Filter: X \times f \mapsto Z$  &{\tt filter, sample, collect}\\    \hline
		{\it Set}        &  $Set: X \times Y \times f \mapsto Z$    & {\tt ++, intersection, union}\\ \hline
		{\it Join}        &  $Join: X \times Y \times f \times K \mapsto Z$    & {\tt join, leftOuterJoin, rightOuterJoin, fullOuterJoin}\\ \hline
		{\it Group}      &  $Group: X \times f \times K \mapsto Z$   & \specialcell{\tt reduceByKey, groupByKey, aggregateByKey, foldByKey}\\ \hline
		{\it Agg}  &  $Agg: X \times f \times init \mapsto reg$       & {\tt reduce, aggregate, fold, max, min}\\
		\hline
	\end{tabular}
	\caption{\small The Definition of Primitive Operations, where $X, Y, Z$ represent datasets and $reg$ is a returned value. $f$ are UDFs working on one or a group of elements. The key $K$ is a subset of attributes shared by two or more datasets, and $init$ is the initial value for aggregate operations. The last column lists representative operations for each category provided by Apache Spark RDD APIs.}
	\label{tab:op_def}
\end{table*}


\begin{itemize}
	\item $Map:X \times f \ \mapsto \{\; f(x) \;| \; x \in X \;\}$ is an operation to return a new dataset by applying $f$ to each element $x$ of $X$. $Flatmap$ is a special map by flattening all elements of the input.
	
	\item $Filter:X \times f \mapsto \{\; x \;| \; x \in X,\;f(x) = True \;\}$ is an operation taking $f$ as a parameter and keeps element $x$ when $f(x)$ is true. $Filter$ reduces the number of elements involved in the successive computation so as to reduce data size for computing and communication later.
	
	\item $Set: X \times Y \times f \mapsto \{\; f(x,y) \;| \; x \in X, \; y \in Y \}$ is an operation on two input datasets, $X$ and $Y$, to generate a new one by applying $f$ to each pair of $\tuple{x,y}$, where the two datasets $X$ and $Y$ should have identical attribute sets.
	
	\item $Join: X \times Y \times f \times K \mapsto \{\; f(x, y) \; | \; x \in X, \; y \in Y, \; x[K] = y[K] \; \}$ is a binary operation on two input datasets, $X$ and $Y$, to generate a new one by applying $f$ to each pair of $\tuple{x,y}$ with matching keys $K$, where $K$ is a subset of attributes shared by both $X$ and $Y$. 
	
	\item $Group: X \times f \times K \mapsto \{\; f(g_{k}) \;| \; g_{k} = \{x_{1},..,x_{m}\} \subseteq X,\;x_1[K]=...=x_k[K]=k\}$ is a unary operation that returns a new dataset by applying $f$ to a group of elements sharing an identical value(s) $k$ on key(s) $K$. 
	
	\item $Agg: X \times f \times init \mapsto f(X,init)$ is to combine all elements in $X$ into a single value with the help of $f$ and an initial value $init$. Note that the result of an operation (\eg, $reduceByKey$) is not a single value, we classify it into a ``Group" operation.
	
\end{itemize}
Actually, there is an implicit {\it Shuffle} operation behind the last four operations to transfer data across processing nodes, which dramatically affects the whole system performance due to expensive I/O operations. One of the ultimate goals of SODA is to reduce the amount of shuffling data as much as possible with the help of our proposed optimization strategies. Although the above definitions just involve at most two input datasets, it is easy to extend the concepts to accommodate more.

\subsection{Data Operational Graph}
\label{subsec:dog}

SODA builds a directed data operational graph (DOG) $G = (V, E)$ to represent an application and conducts three kinds of optimizing strategies atop this graph. A vertex $v \in V$ depicts a primitive operation described in Table \ref{tab:op_def} and the dataset generated by this operation. An edge $e \in E$ denotes data flows between two operations. For each vertex, there is a group of properties accumulated from static analysis, dynamic analysis, or both on code, data and runtime system, which is defined in Table~\ref{table:stats} in more detail. We also add two special nodes, named as {\it Source} and {\it Sink}. {\it Source} node is connected to all initial input datasets while all sole output of stages would point to {\it Sink} node. SODA conducts optimizations atop of a DOG, rather than on an abstract syntax tree (AST) due to the following considerations: 1) usually a data-intensive system supports various program language APIs (\eg, Scala, Java, python APIs in Apache Spark), a general optimization backend is compatible with different programming models; 2) SODA focuses on optimizations at the level of operations, rather than at the lower level of AST nodes; 3) There is a huge gap between AST nodes and simulating system behaviors that interpret applications and datasets.



\begin{figure}[ht!]
	\centering
	\includegraphics[scale=0.3]{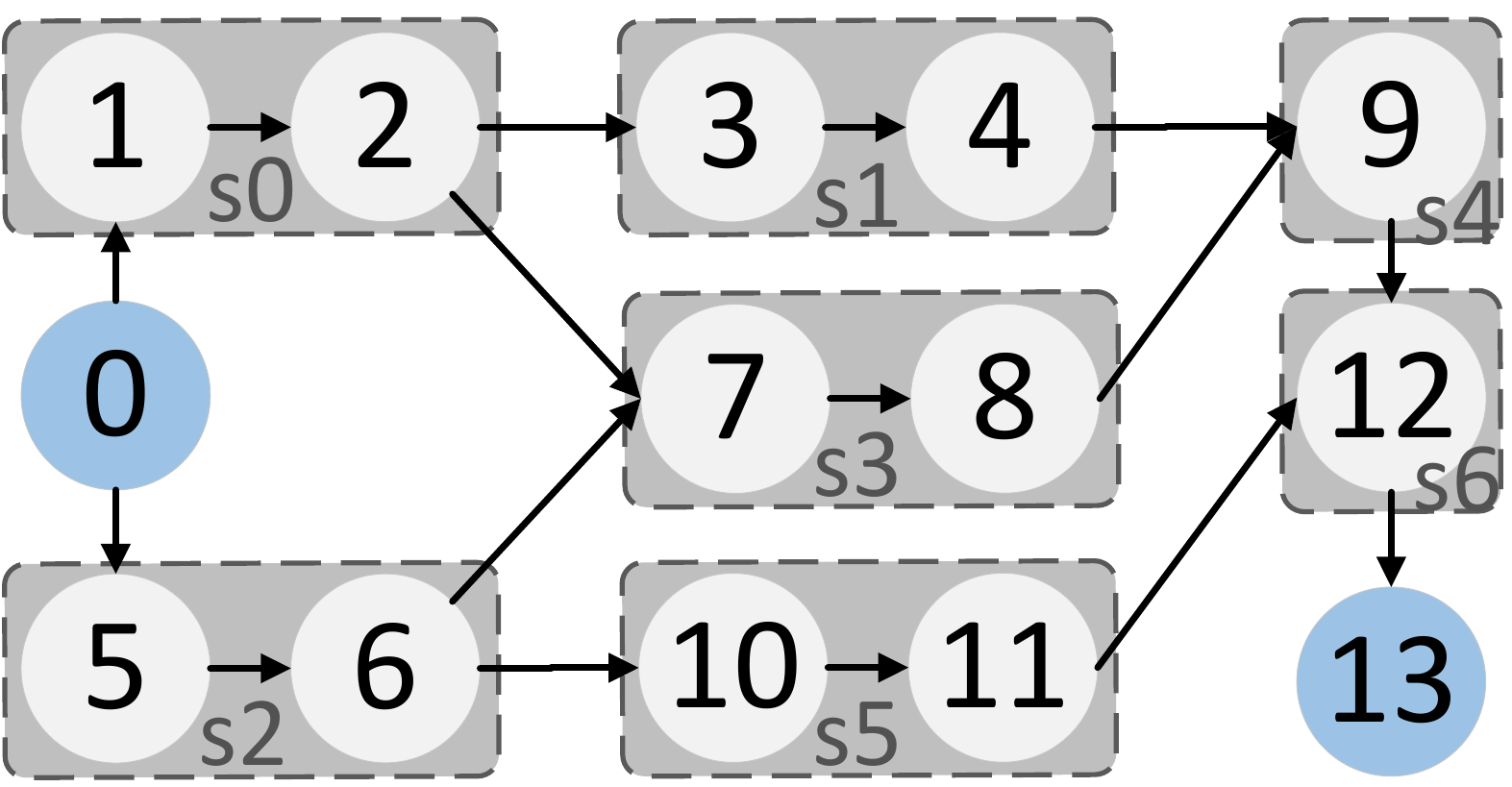}
	\caption{Data Operational Graph of Customer Reviews Analysis benchmark, with stages~(wrapped by dashed rectangles and labeled by texts starting with s) and data blocks with dependencies~(indicated by solid arrows).}
	\vspace{-4mm}
	\label{fig:dog_cache_sub}
\end{figure}

{\bf Execution model}. Without loss of generality, SODA splits an execution plan of DOG into a series of stages that are bounded by shuffle behaviors, denoted by $S = \{s_1,s_2,\ldots,s_n\}$. As shown in Figure~\ref{fig:dog_cache_sub}, the toy application is composed of seven stages. A stage $s \in S$ is delegated as a physical scheduling unit consisting of multiple operations to fulfill a sub-job. Generally, a stage $s$ involves an execution path between the {\it Source} node and its target $v_t$~(\ie, $s.target$) if no cache mechanism is provided: $s = \{v_0,\ldots, v_t\}$. For example, ${s_3} = \{v_0, v_1, v_2, v_5, v_6, v_7, v_8\}$ is a set of nodes involved in computing the outcome~(\ie\;$v_{8}$) of stage $s_3$ in Figure~\ref{fig:dog_cache_sub}. Generally speaking, the computational cost of a stage $s$ is calculated by aggregating all involved operations' execution time, denoted as $C_{s} = \sum_{v \in s}T_{v}$ where let $T_v$ denote the execution time of an operation of node $v$. Furthermore, The total execution time of an application is given by summing all stages' cost: $C_{S} = \sum_{s \in S} C_s$. 

It is well known that stages can run in parallel if there is no data dependency among them in a data-intensive system. However, without loss of generality, we assume that they are scheduled in sequential order. SODA determines this order by analyzing the data dependency of stages and the submission time $T_s$ of stages in prior executions extracted from the performance log. An operation can be executed 
simultaneously by a cluster of executors on different data partitions. Technically, these executors can be equipped with the configurable size of computing resource (\eg, CPU, memory). We also assume that memory resource in an executor is divided into two sections for storage (\ie, caching intermediate data) and computation(\ie, allocating objects). We denote $M_{store}$ as the size of storage memory.

\section{Optimization Strategies}
\label{sec:opt}

There are three kinds of optimization strategies: cache management, operation reordering and element pruning.

\subsection{Cache Management}
\label{subsec:perf_mm}

In this section, we go over the details about Cache Management policy. The summary notation is categorized and listed in Table~\ref{table:stats}.

{\bf Maximizing Expected Caching Gain.} A global cache allocation is usually preferred to minimize the aggregated execution cost of an application. In particular, we assume $C_0$ is the real executing time of an application without any optimizations and works as an upper bound on the expected costs. Here our objective is to determine a feasible cache allocation~(\ie\; $w$) that maximizes the caching gain, \ie, the expected cost reduction attained by caching data, which is defined as: $F(w) = C_0 - \sum_{s \in S}C^{'}_s$, where $C_{s}'$ is defined as the predicted (or expected) computational cost of a stage $s$ by consideration of $w$. 


To determine a global cache allocation policy, a binary matrix $W=[w_{sv}]_{s \in \mathcal{E}_S, v \in V} \in \{0,1\}^{|\mathcal{E}_S|\times|V|}$ is defined to indicate cache status of a data generated by node $v$ after executing a stage $s$, where $\mathcal{E}_{S} = \{\mathcal{E}_{s_1},\mathcal{E}_{s_2}, \dots, \mathcal{E}_{s_n}\}$ reveals the real-time scheduling order of all stages extracted from online profiling information. In the matrix, a cell with value 1 (\ie, $W[s,v] = 1$) indicates that the output of the data of $v$ is reserved in memory after a stage $s$ is done (See  Equation~\ref{subeq:matrix_cache}); otherwise, \ie, when $W[s,v] = 0$, the data is evicted from memory (See  Equation~\ref{subeq:matrix_uncache}). It is worth mentioning that cache capacity constraints in an executor~($M_{store}$ is the size of memory for storage) would limit the amount of involved data that could be reserved in memory (See  Equation~\ref{subeq:matrix_capacity}). From top to bottom in a column of $W$, it is easy to identify which stage a data is stored into memory, and which stage it is evicted from memory. Such an allocation plan tells programmers when to persist or unpersist data in memory in code. 


Given a global cache allocation, all operations involved in the computation of a stage $s$ are well routed by following the execution path until it encounters a data of $v$ cached in memory. This data and its predecessors do not need to be recomputed so far. In the previous example of $C_{s_3}$, the cost is equal to $T_{v_7} + T_{v_8}$ if data generated by $v_2$ and $v_6$ are cached in memory. Next, given the current executing stage $s$ with a global cache allocation $w \in W$, the number of re-computation times of $v_k \in V$ (because it is used again later but not cached) is needed to get the outcome of $v_l \in V$, which defined in Equation~(\ref{eq:cache_prop}).

\begin{small}\begin{equation}
\label{eq:cache_prop}
P(v_k,v_l, s) = \sum_{p \in \tau(v_k,v_l)} \prod_{v \in p}(1 - w[s.pred, v]))
\vspace{-2mm}
\end{equation}\end{small}where $\tau(v_k, v_l)$ returns a set of paths from node $v_k$ to $v_l$; if $v_k$ is identical to $v_l$, then it is $\{\{v_k\}\}$; $s.pred$ reveals the previous executing stage of $s$. Therefore, the predicted (or expected) computational cost of a stage $s$ can be regulated concisely under a global cache allocation policy $w \in W$, and defined in Equation~(\ref{eq:cost_stage}).

\begin{small}\begin{equation}
\label{eq:cost_stage}
\vspace{-2mm}
\begin{split}
C_{s}' &= \sum_{v \in s}^{v_t = s.target} T_v * P(v,v_t,s) \\
&= \sum_{v \in s}^{v_t = s.target} T_v * \sum_{p \in \tau(v,v_t)} \prod_{v' \in p}(1 - w[s.pred, v']))
\end{split}
\vspace{-2mm}
\end{equation}\end{small}

Finally, we try to obtain an allocation of policy $w$ that maximizes the aggregate expected caching gain:

\begin{small}\begin{equation}
\vspace{-2mm}
\label{eq:exp_obj_f}
\begin{split}
F(w) &= \sum_{s \in S}C_s - \sum_{s \in S}C_s' \\
&= C_0 - \sum_{s \in S}\sum_{v \in s}^{v_t = s.target} T_v * \sum_{p \in \tau(v,v_t)} \prod_{v' \in p}(1 - w[s.pred, v']))
\end{split}
\vspace{-2mm}
\end{equation}\end{small}


{\bf Convex-Concave Relaxation.} In particular, we seek solutions to the following problem:

\begin{small}\begin{subequations}
	\label{eq:fx}
	\begin{align}
	\argmax F(w) \label{eq:fx_obj}\\
	\subjectto \; w \in D_{1} \label{eq:fx_constrait}
	\end{align}
\end{subequations}\end{small}where $D_1$ is the set of matrices $W \in \{0, 1\}^{|\mathcal{E}_S|\times|V|}$ satisfying source constraints, cache behaviors and cache capacity, \ie,

\begin{small}\begin{subequations}
    \vspace{-2mm}
	\label{eq:matrix_opt}
	\begin{align}
	\forall s \in \mathcal{E}_S, v \in V : W[s,v] \in \{0, 1\}     \label{subeq:matrix_value}\\
	\exists s \in \mathcal{E}_S, v \in V, W[s, v] = 1: s \xrightarrow{cached} v\label{subeq:matrix_cache}\\
	\exists s \in \mathcal{E}_S, v \in V, W[s, v] = 0: s \xrightarrow{uncached} v \label{subeq:matrix_uncache}\\
	\forall s \in \mathcal{E}_S:\sum_{v \in V}W[s,v] * S_{v} \leq M_{store} \label{subeq:matrix_capacity}
	\vspace{-2mm}
	\end{align}
\end{subequations}\end{small}where $S_v$ denotes the size of an intermediate data generated by an operation of $v$. As far as we know, this deterministic, combinatorial version of~(\ref{eq:fx}) is NP-hard, even when we already have background knowledge about the submitted application and runtime statistics. Nonetheless, we can relax it to a submodular maximization problem subject to knapsack constraints and take linear relaxation algorithm to optimize cache allocation on the stage level by minimizing the expected computational cost~\cite{ioannidis2016adaptive, yang2018intermediate}. It is obvious that Equation~(\ref{eq:fx}) is not a convex optimization problem. However, it can be approximated as follows. We can define $L: W \rightarrow \field{R}$ based on Equation~(\ref{eq:exp_obj_f}) as:

\begin{small}\begin{equation}
\vspace{-2mm}
\begin{split}
L(w) &= C_0\;- \\
&\sum_{s \in S}\sum_{v \in s}^{v_t = s.target} T_v *\sum_{p \in \tau(v,v_t)}( 1- min(1,\sum_{v' \in p}w[s.pred, v']))
\end{split}   
\vspace{-2mm}
\end{equation}\end{small}Note that $L$ is a concave function, and now we have the following: 
\begin{small}\begin{subequations}
	\label{eq:lx}
	\begin{align}
	\argmax L(w) \label{eq:lx_obj}\\
	\subjectto \; w \in D_1 \label{eq:lx_constrait}
	\end{align}
\end{subequations}\end{small}
According to~\cite{ioannidis2016adaptive}, an optimal solution $w$ to~(\ref{eq:lx}) can be approximated and guaranteed within a constant factor $(1 - 1/e)$ from the optimal value of Equation~(\ref{eq:fx}): $(1 - 1/e)L(w) \leq F(w) \leq L(w), \forall w \in D_1$.

{\bf Global Execution Distance.} So far, SODA can approximate a solution to~(\ref{eq:lx}) within a $(1 - 1/e)$ factor by searching all cache allocation space, which may lead to a bad runtime performance. In other words, We convince that knowledge about data flow and stages' dependency could have a positive effect on this defect. Therefore, we devise a new metric to measure the time-locality distance of an operation, namely execution distance, and introduce another constraint to $D_1$. 


\begin{definition}[{\bf Global Execution Distance~(GED)}]
	\label{def:ed}
	For a node $v \in V$, execution distance is defined as a relative difference between the current execution point $\mathcal{S}_c$ and a future executing stage $\mathcal{S}_f$ in which it will be referenced: $\mathcal{S}_f - \mathcal{S}_c$.
\end{definition}

In particular, there may have multiple execution distances for the data of $v$ if it is used in several stages. At this point, the final number should be the sum of all these distances. For instance, Table~\ref{tab:dog_cache_table} shows an evolution of execution distance for each node in Figure~\ref{fig:dog_cache_sub} as the workload runs along with scheduling order $\mathcal{E}_S$ from top to bottom. In the first row of the table, we have twelve operations, which may be cached in memory after a stage is done; The leftmost two columns reveal the relationship between stages $S$ and their corresponding scheduling order $\mathcal{E}_S$. The number in the rest of the cells indicates how far away from a future reference point to the current executing stage, and it should be recalculated and updated after each execution of stages every time. For example, after executing stage $s_2$~(its corresponding schedule order is 1), the execution distance of $v_2$ is updated from 5 to 3 since $v_2$ will be referred in stage $s_1$ and $s_3$ and their corresponding schedule order is 2 and 3, respectively. So the new value will be recalculated by $(2 - 1) + (3 - 1)$. A cell~$[s,v]$ can be set to zero if 1) the data generated by $v$ is referenced by another node in the same stage $s$~(See case cell of $[0,v_1]$); 2) the data of $v$ gets referenced and there is no more reference in the future~(See case cell of $[3,v_2]$). The cells with empty content mean the nodes that have not been accessed so far.

\begin{table}
	\centering
	\begin{minipage}[t]{1.\linewidth}
		\resizebox{\linewidth}{!}{%
			\begin{tabular}{|c|c|c|c|c|c|c|c|c|c|c|c|c|c|} \hline
				$\mathcal{E}_S$ &$S$ &$v_1$ &$v_2$ &$v_3$ &$v_4$ &$v_5$ &$v_6$  &$v_7$ &$v_8$ &$v_9$ &$v_{10}$ &$v_{11}$ &$v_{12}$\\\hline
				0 &$s_0$    &0 &5 &  & & & & & & & & &\\\hline
				1 &$s_2$    &0 &3 &0 &0 &0 &6 & & & & & &\\\hline
				2 &$s_1$    &0 &1 &0 &2 &0 &4 & & & & & &\\\hline
				3 &$s_3$    &0 &0 &0 &1 &0 &2 &0 &1 & & & &\\\hline
				4 &$s_4$    &0 &0 &0 &0 &0 &1 &0 &0 &2 & & &\\\hline
				5 &$s_{5}$  &0 &0 &0 &0 &0 &0 &0 &0 &1 &0 &1 &\\\hline
				6 &$s_{6}$  &0 &0 &0 &0 &0 &0 &0 &0 &0 &0 &0 &0\\\hline
			\end{tabular}
		}%
	\end{minipage}
	\caption{The cache allocation policy based on Execution Distance for the workload in Figure~\ref{fig:dog_cache_sub}}
    \vspace{-8mm}
	\label{tab:dog_cache_table}
\end{table}


With the help of GED, we can also learn a set of candidates that can be persisted in memory after a stage $s$ is finished, termed as $\mathcal{H}_s$. For example, $\mathcal{H}_{s_1} = \{v_2, v_4, v_6\}$ since the corresponding cells are non-zero in the row of $\mathcal{E}_S(=2)$. Therefore we can narrow down search space to approach an optimal solution to~(\ref{eq:lx_opt}) by merely considering data in $\mathcal{H}_S$, rather than all data in $V$, for a stage $s$. Consider the following problem:

\begin{small}\begin{subequations}
	\label{eq:lx_opt}
	\begin{align}
	\argmax L(w) \label{eq:lx_opt_obj}\\
	\subjectto \; w \in D_{2} \label{eq:lx_opt_constrait}
	\end{align}
\end{subequations}\end{small}
where $D_2$ is the set of matrices $W \in \{0, 1\}^{|\mathcal{E}_S|\times|V|}$ satisfying source constraints, cache behaviors, cache capacity, and hypothesis of $\mathcal{H}_s$, \ie:

\begin{small}\begin{subequations}
    \vspace{-2mm}
	\label{eq:matrix}
	\begin{align}
	\forall s \in \mathcal{E_S}, v \in V : W[s,v] \in \{0, 1\}     \label{subeq:opt_matrix_value}\\
	\exists s \in \mathcal{E_S}, v \in V, W[s, v] = 1: s \xrightarrow{cached} v\label{subeq:opt_matrix_cache}\\
	\exists s \in \mathcal{E_S}, v \in V, W[s, v] = 0: s \xrightarrow{uncached} v \label{subeq:opt_matrix_uncache}\\
	\forall s \in \mathcal{E_S}:\sum_{v \in V}W[s,v] * S_{v} \leq M_{store} \\
	\forall s \in \mathcal{E_S}, v \in (V \setminus \mathcal{H}_s): W[s, v] = 0
	\label{subeq:opt_matrix_capacity}
	\end{align}
	\vspace{-2mm}
\end{subequations}\end{small}

It is apparent that $D_2$ is a subset of $D_1$, a solution $w'$ to~(\ref{eq:lx_opt}) can also be fit for~(\ref{eq:lx}), as well as~(\ref{eq:fx}) with $(1 - 1/e)L(w') \leq F(w') \leq L(w'), \forall w' \in D_2$. To gain better approximating rate, we implement Pipage Rounding~\cite{ioannidis2016adaptive} using Gurobi optimizer APIs~\cite{optimization2014inc} to approximate a solution to~(\ref{eq:lx_opt}).

\subsection{Operation Reordering}
\label{subsec:perf_om}

The goal of operation reordering~(\ie\; Filter Pushdown) is to improve applications' performance by reordering operations along with data path. There are two challenges: Is reordering correct concerning the original semantics? Does the reordering improve performance? To answer these questions, we first define Use-Set and Def-Use by following the dataflow technique in static code analysis~\cite{nielson2015principles}.

\begin{definition}[{\bf Use-Set}]
\label{def:u_op}
	Given $Y$=$X.op(f)$, Use-Set $U_{f} = \{a~|~a \in \beta(X)$ and $a$ is accessed by $f$\}. Use-Set defines all attributes of input data used by $f$ to generate $Y$.
\end{definition}


\begin{definition}[{\bf Def-Set}]
	\label{def:d_f}
	Given $Y = X.op(f)$, Def-Set $D_{f} = \{b~|~b \in \beta(Y)$ and b is created or updated by $f$\}. Def-Set is the attribute set newly created by an operation $op$, or inherited directly from $\beta(X)$.
	
\end{definition}

Then, SODA uses a two-step way to handle these two challenges. In the first step (static verification), Theorem~\ref{theory:reodering} is proposed to ensure semantic correctness. It captures the fact that two successive operations can be reordered if a latter UDF $f_2$ does not use attributes that a former UDF $f_1$ defines. 

\begin{theorem}
	\label{theory:reodering}
	Two successive operations $op_1$ and $op_2$ on an execution path can be reordered, \ie, $X.op_1(f_1).op_2(f_2) \equiv X.op_2(f_2).op_1(f_1)$, if $U_{f_2} \cap D_{f_1} = \emptyset$.
\end{theorem}

Let's take filter pushdown as an example to illustrate this theorem. Filter pushdown is a conventional optimization that pushes a {\it filter} towards the direction of data loading as much as possible so that the volume of intermediate data can be reduced. 

\begin{lemma}
\label{lemma:filter_map} 
For $Y=X.Map(f_1).Filter(f_2)$, $Filter$ and $Map$ can be reordered, if $U_{f_2} \cap D_{f_1} = \emptyset$.
\end{lemma}

The comprehensive proof statement of Lemma \ref{lemma:filter_map} is followed: 

\vspace{-8pt}
\begin{proof} Assume the two plans 
$$O^{1} = X.Map(m).Filter(f)$$
$$O^{2} = X.Filter(f).Map(m)$$
We prove that $O^1\equiv O^2$. Assume a record $x \in X$, let $O_x^1=f(m(x))$ and $O_x^2=m(f(x))$ are set of element(s) generated by applying according operations by sequence to $x$. Notice that here $O^{1}= \bigcup\limits_{x\in X} f(m(x))$, and $O^{2} = \bigcup\limits_{x\in X} m(f(x))$, and in all our proofs, set is referring to dataset (mathematically termed as a multiset) which allows repetitive elements and union operations (here alias to sum operation in multiset) preserve repetitive elements as well. To prove $O^1\equiv O^2$, it suffices to show that $\forall x \in X : O_x^1\equiv O_x^2$. We prove it by justifying the following two cases: 
$1_f(x)$ is the indicator function of filter $f$ to represent its selectiveness.
\textbf{1.} When $f(x)=1_f(x)\cdot x=0\cdot x=\emptyset$, where  Then, $O_x^2=m(f(x))=m(\emptyset)=\emptyset$. Now since $U_{Filter} \cap D_{Map} = \emptyset$, we know that for $\forall x'\in m(x)$, $\pi_{U_{Filter}}(x') = \pi_{U_{Filter}}(x)$, and by definition~\ref{def:u_op}, $f$'s behavior is solely depending on attribute set $U_{Filter}$, we have
\begin{align*}
    O_x^1&=f(m(x))=\bigcup\limits_{x'\in m(x)} f(x')=\bigcup\limits_{x'\in m(x)} 1_f(x')\cdot x'\\
    &=\bigcup\limits_{x'\in m(x)} 1_f(\pi_{U_{Filter}}(x'))\cdot x' =\bigcup\limits_{x'\in m(x)} 1_f(\pi_{U_{Filter}}(x))\cdot x'\\
    &=\bigcup\limits_{x'\in m(x)} 1_f(x)\cdot x'=\bigcup\limits_{x'\in m(x)} 0\cdot x'=\emptyset
\end{align*}

\noindent Thus for $\forall x$, in case 1, we have $O_x^1\equiv O_x^2$.
\vspace{2mm}

\textbf{2.} Similarly, when $f(x)=1_f(x)\cdot x=1\cdot x=\{x\}$, $O_x^2=m(f(x))=m(x)$. And
\begin{align*}
    O_x^1&=f(m(x))=\bigcup\limits_{x'\in m(x)} f(x')=\bigcup\limits_{x'\in m(x)} 1_f(x')\cdot x'\\
    &=\bigcup\limits_{x'\in m(x)} 1_f(\pi_{U_{Filter}}(x'))\cdot x'=\bigcup\limits_{x'\in m(x)} 1_f(\pi_{U_{Filter}}(x))\cdot x'\\
    &=\bigcup\limits_{x'\in m(x)} 1_f(x)\cdot x'=\bigcup\limits_{x'\in m(x)} 1\cdot x'=\bigcup\limits_{x'\in m(x)} x'=m(x)
\end{align*}

\noindent Thus for $\forall x$, in case 2, we also have $O_x^1 \equiv O_x^2$. Combine the results in both cases (which are all cases possible), we have proved for $\forall x\in X$, $O_x^1\equiv O_x^2$, and consequently, $O^1\equiv O^2$. 
\end{proof}

Correspondingly, we can get the following lemmas to determine if a {\it Filter} operation can be pushed down before {\it Group} and {\it Set} operations, respectively. 

\begin{lemma}
\label{lemma:filter_group}
For $Y=X.Group(f_1).Filter(f_2)$, $Filter$ and $Group$ can be reordered, if $U_{f_2} \cap D_{f_1} = \emptyset$.
\end{lemma}

\begin{lemma}
\label{lemma:filter_set}
For $Z=X.Set(Y, f_1).Filter(f_2)$, $Filter$ and $Set$ can be reordered along with $X$ and $Y$ data path safely: \\  
$Z=X.Filter(f_2).Set(f_1)(Y.Filter(f_2))$ if $U_{f_2} \cap D_{f_1} = \emptyset$.
\end{lemma}





In the second step~(dynamic evaluation), two polynomial regression models~(due to their wide applicability in engineering~\cite{gibilisco2016stage}) are trained for $op_1$ and $op_2$ respectively using profiling information, then predict the execution time of each operation on new input. If SODA gets positive feedback from predict models, it will suggest programmers reorder these two operations.

\subsection{Element Pruning}
\label{subsec:perf_rp}

Element pruning is an optimization to eliminate unused attributes in an element by analyzing data dependency in the attribute level among operations. SODA analyzes an operation and its associated UDF(s) to analyze attribute dependency between the input and output dataset of this operation. Then a directed data dependency graph (DDG), $G' = (V', E')$, is built to represent the whole data flow of the application by combining all attribute dependency relationships among operations. A node $v \in V'$ represents an attribute of an dataset involved in an operation while an edge $e \in E'$ from a node {\it s} to another node {\it d} indicates that {\it d} has either data or control dependency on {\it s}. If an edge is a control dependency, it means that {\it s} and {\it d} have identical attributes. 
Data dependency means that the value of {\it d} is updated or created from {\it s}. An attribute node may have multiple incoming and outgoing edges. 
To identify an application's start and endpoints, we add two special nodes {\it source} and {\it sink} to this graph and connect all input attributes of this application to {\it source} and connect all output attributes to {\it sink}. All these dummy edges outgoing from {\it source} and incoming to {\it sink} are assigned as control dependencies. Therefore, we can reduce the complicated optimization into a problem of traversing the graph and eliminating a node $v$ if there exists no path between $v$ and {\it sink}, since an attribute node can be eliminated safely if it does not make a contribution to produce an output of the application. 

\begin{figure}[ht!]
    \vspace{-2mm}
	\centering
	\includegraphics[scale=0.6]{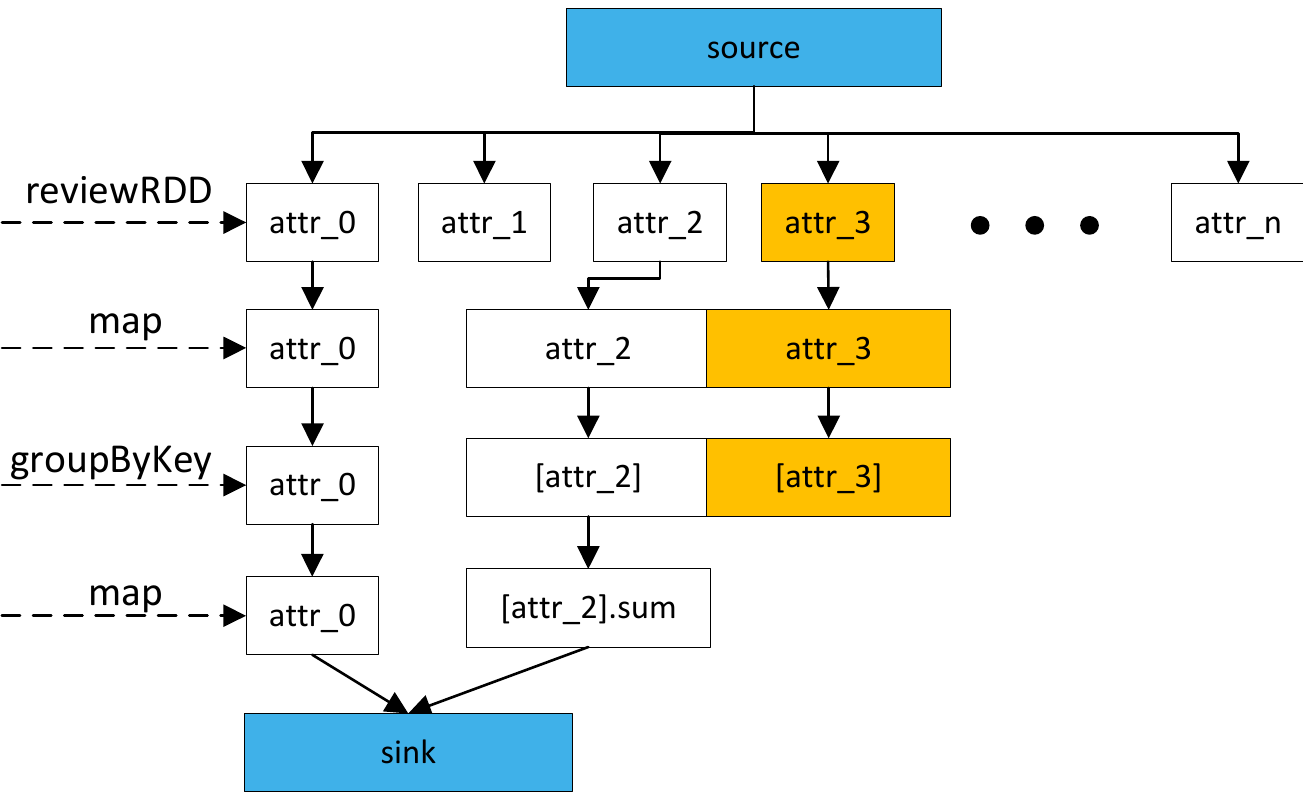}
	\caption{A simplified example of data dependency tree}
		\vspace{-2mm}
	\label{fig:data_dep_graph}
\end{figure}

Figure~\ref{fig:data_dep_graph} shows an example of a data dependency graph of Listing~\ref{lst:unusedelement}. Each row represents a group of attributes of a dataset named by the corresponding leftmost text above the dashed arrow. A rectangle reveals an attribute labeled by the inside text. It is obvious that the attribute ``[attr\_3]" does not contribute to ``sink" while it is grouped by {\it groupByKey} operation from the attribute ``attr\_3" in first {\it map}. The preliminary experiment shows that this kind of awkward design leads to a significant computation and I/O cost because of shuffling a huge size of data among computing nodes over the network. According to our proposed constraint, there is no edge between these yellow rectangles and ``sink"  so they can be removed without changing the snippet code purpose.

\vspace{-3mm}
\begin{lstlisting}[caption={An example showing the problem of EP}, label={lst:unusedelement},captionpos=t]
val aggData = reviewRDD.map( row <@$\Rightarrow$@>
 (row.getString(0),(row.getDouble(2),<@\textcolor{red}{row.getString(3)}@>))
).groupByKey().map{
case (attr_0,attr_2) <@$\Rightarrow$@> attr_2.map(_._1).sum }
\end{lstlisting}
\vspace{-1mm}

\section{Evaluation \& Experiments }
\label{sec:exper} 


In this section, we use four real-world data-intensive applications in different domains to evaluate the overall effectiveness of SODA on a 9-node cluster of Apache Spark (v3.0.0) by comparing runtime performance of these applications before and after optimization by SODA. Each node has a hardware configuration with Intel(R) Xeon(R) CPU E5-2620 v3 @ 2.40GHz, 32GB main memory with DDR4-2133 MHz ECC and 1 GigE Ethernet as the internal communication channel between nodes.



\begin{table*}[ht!]
	\centering
	\begin{minipage}[t]{0.7\linewidth}
		\resizebox{\linewidth}{!}{%
			\begin{tabular}[4]{c|c|l|l}
				\hline
				Level   &Notation    &Data Source   &Comments\\
				\hline
				\multirow{7}{*}{Application}
				& $G=(V, E)$  & Source code & Data Operational Graph with nodes V and edges E\\ \cline{2-4}
				& $S$  & Source code, System Log & All stages in an application\\ \cline{2-4}
				& $W$  & Defined by SODA & A binary matrix $W \in \{0, 1\}^{|\mathcal{E}_S|\times|V|}$\\ \cline{2-4}
				& $D_1$  & Defined by SODA & Set of matrices $W \in \{0, 1\}^{|\mathcal{E}_S|\times|V|}$\\ \cline{2-4}
				& $D_2$  & Defined by SODA & Set of matrices $W \in \{0, 1\}^{|\mathcal{E}_S|\times|V|}$ satisfying hypothesis of $\mathcal{H}_s$\\ \cline{2-4}
				& $F(w)$  & Defined by SODA & The expected caching gain~(\ref{eq:fx}) in $D_1$\\ \cline{2-4}
				& $L(w)$  & Defined by SODA & The concave approximation~(\ref{eq:lx}) of $F(w)$\\ \hline
				
				\multirow{3}{*}{Stage}
				& $s$  & Source code, System Log & A stage in an application\\ \cline{2-4}
				& $\mathcal{H}_s$  & Defined by SODA & Cache candidate datasets after a stage $s$ is finished\\ \cline{2-4}
				& $C_s$  & Defined by SODA & The computational cost of a stage $s$\\ \cline{2-4}
				& $T_s$  & System Log, Runtime Profier & The submission time of a stage $s$\\ \hline
				
				\multirow{3}{*}{Operation}
				& $T_v$  & System Log, Runtime Profier & The execution time of an operation of $v$\\ \cline{2-4}
				& $U_{f}$  & Source code & A Use-Set of an operation $op$ with UDF $f$\\ \cline{2-4}
				& $D_{f}$  & Source code & A Def-Set of an operation $op$ with UDF $f$\\ \cline{2-4}
				
				\hline
				\multirow{2}{*}{Dataset}
				& $S_v$ & System Log, Runtime Profier & The size of a dataset generated by operation $v$ \\ \cline{2-4}
				& $N_{v}$ & System Log, Runtime Profier & The number of elements in a dataset generated by operation $v$\\ \hline
				
				\multirow{2}{*}{System}
				&$M_{exe}$   & System Log & The memory size of an executor\\ \cline{2-4}
				&$M_{store}$ & Defined by SODA & The size of storage memory \\ \cline{2-4}
				\hline
			\end{tabular}
		}%
	\end{minipage}
	\caption{\small The statistics information and  corresponding notations needed by SODA}
	\vspace{-4mm}
	\label{table:stats}
\end{table*}

\subsection{Benchmarks}
\label{subsec:workloads}

\begin{itemize}
    
    \item {\bf System Log Analysis (SLA)} is a job to find average ranking and total advertising revenue for each website within a specified date range. There are two datasets, uservisits and pageranks. 

    \item {\bf Customer Reviews Analysis (CRA)} is a project aiming at ranking the top 20 brands according to average customer rating score in the book categories. The review datasets include over 138.1 million customer reviews spanning from May 1996 to July 2014~\cite{mcauley2016addressing}.


    \item {\bf Social Network Analysis (SNA)} focuses on ranking the top 20 users who are the most active in a specified time period based on tweets analysis. We use a social-media community consisting of 790,462 users who posted over 3,286,473 tweets and have more than 3,055,797 links from 2013 to 2015~\cite{chung2017itsec, chung2019interaction}.


    \item {\bf Pre-Processing Job (PPJ)} is a clean task and looks for products satisfying two criteria: 1) product ID starts with ``B000"; 2) average word count of the product description is greater than 100. N/A data element will be removed to avoid program crashes during runtime. The metadata dataset includes 15.5 million products.  
\end{itemize}



\subsection{Effectiveness Assessment}
\label{subsec:funcAssess}

\begin{table}[ht!]
	\centering
	\begin{minipage}[t]{1\linewidth}
		\resizebox{\linewidth}{!}{%
		    \begin{tabular}{|c|l|c|c|c|}
         \hline 
         Bechmark &Description & CM & OR & EP \\ \hline
         SLA&Filter, Join, Agg & Detected      &Not Present &Detected\\ \hline
         CRA&Filter, Join, Agg&Detected &Detected &Detected\\ \hline
         SNA&Map, Filter, Agg&{\it Failed} &Detected &Detected\\ \hline
         PPJ&Map, Filter, Group&Detected &Not Present &Detected\\ \hline
    \end{tabular}
		}%
	\end{minipage}
\caption{\small The results of running SODA on Spark Applications. CM, OR and EP represent Cache Management, Operation Reordering and Element Pruning, respectively.}
    \label{tab:effectness}
	\vspace{-3mm}
\end{table}


To evaluate the effectiveness of SODA, we first manually examine all source code to see which problems are {\it present} by rules-of-thumb. We then apply each optimization on four benchmarks individually to obtain their results in detecting problems: {\it Detected}, {\it Undetected}, or {\it Not Present}.  If a problem is detected but the performance behaves worse after the revision, we label it as a {\it Failed} case. The results shown in Table~\ref{tab:effectness} allow for quantifying their performance. In general, most potential performance problems are detected by SODA successfully with one exception of a {\it Failed} case in SNA workload when being applied with CM optimization.


\begin{itemize}
    \item {\bf SLA} is an application working on two datasets to evaluate the performance of CM and EP, and OR is not applied in this application. SODA can scrutinize the problems successfully.
    
    \item {\bf CRA} is a complicated student project using Filter, Join, Agg operations, which exposes problems of EP and OR to SODA. In addition, the complicated workflow allows SODA to dig into the CM issue. All of the problems can be detected by SODA successfully. 
    
    \item {\bf SNA} is a research project that involved all the optimizations. All of them can be detected by SODA statically, however, CM leads to negative feedback regarding execution time while the other two have positive effects on the application. We, therefore, label CM as {\it Failed}. More discussion regarding this unusual phenomenon will be given in the next section.
    
    
    \item {\bf PPJ} is a data clean task involved in Map, Filter and Group operations. There are two problems of CM and EP, are successfully detected by SODA.
\end{itemize}

\subsection{Performance Behavior}
\label{subsec:Indi_Perf}

\begin{figure*}[ht!]
	\centering
	\begin{subfigure}[t]{0.25\textwidth}
		\centering
		\includegraphics[scale=0.35]{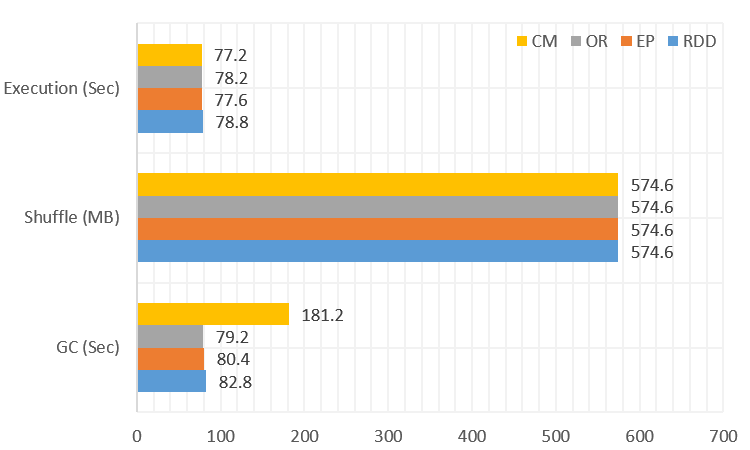}
		\caption{SLA}
		\label{subFigure:perf_sla}
	\end{subfigure}%
	~ 
	\begin{subfigure}[t]{0.25\textwidth}
		\centering
		\includegraphics[scale=0.35]{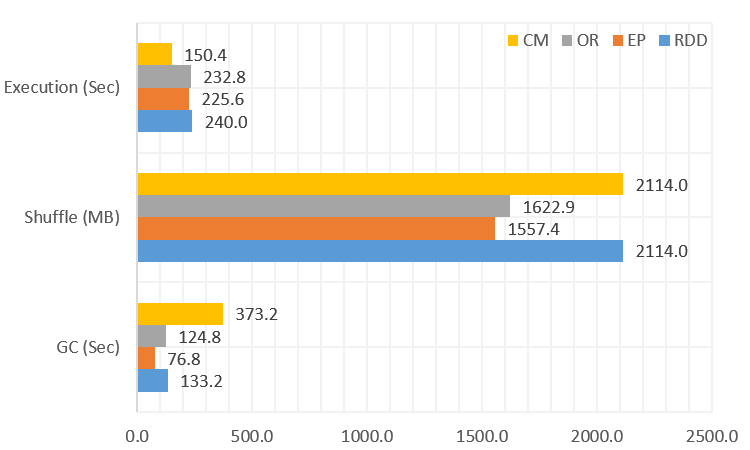}
		\caption{CRA}
		\label{subFigure:perf_cra}
	\end{subfigure}%
	~
	\begin{subfigure}[t]{0.25\textwidth}
		\centering
		\includegraphics[scale=0.35]{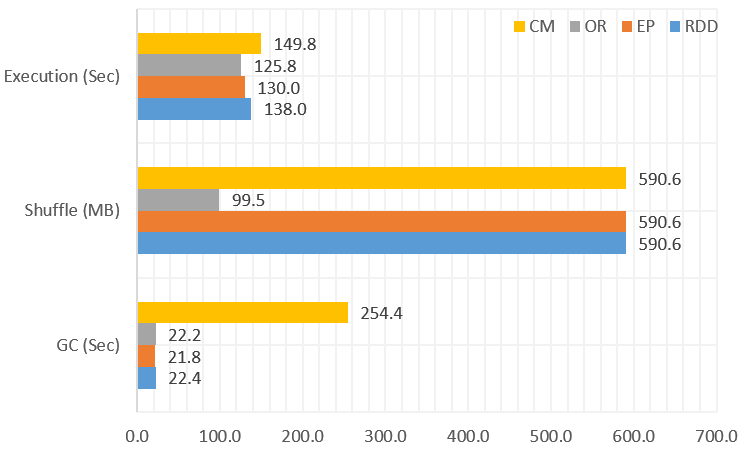}
		\caption{SNA}
		\label{subFigure:perf_sna}
	\end{subfigure}%
	~ 
	\begin{subfigure}[t]{0.25\textwidth}
		\centering
		\includegraphics[scale=0.35]{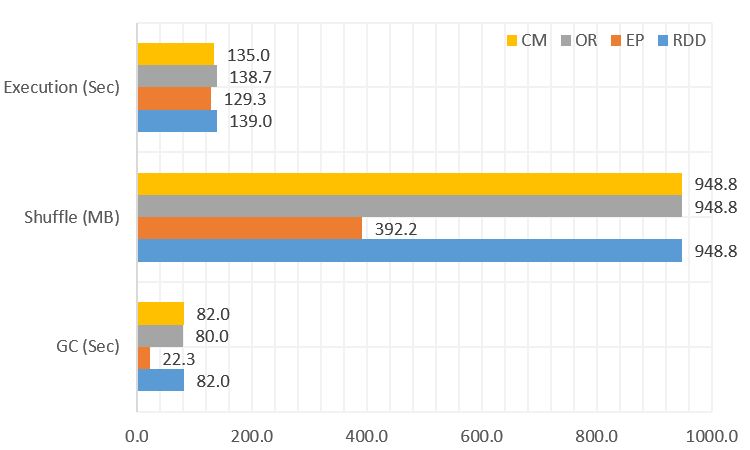}
		\caption{PPJ}
		\label{subFigure:perf_ppl}
	\end{subfigure}
	\caption{\small The performance of individual optimization over the baseline. Label ``RDD" refers to the performance of baseline without any optimizations. Labels ``CM", ``OR", ``EP" represent performances of applications optimized by cache management, operation reordering and element pruning, respectively. Each describes experimental results in terms of execution time, shuffling data size and garbage collection time.}
		\vspace{-4mm}
	\label{figure:Perf_individual}
\end{figure*}

We start the performance improvement evaluation of SODA on workloads for each optimization. We implemented this by submitting the revised code to spark and run it five times for each workload to obtain average experimental data. Figure~\ref{figure:Perf_individual} shows the experimental results of execution time, size of shuffling data and GC time for each benchmark. Table~\ref{tab:perf_speedup} lists the speed up each optimization achieves for each benchmark.  
 
\begin{table}[ht!]
	\centering
	\begin{minipage}[t]{0.7\linewidth}
		\resizebox{\linewidth}{!}{%
		   \begin{tabular}{|c|c|c|c|}
         \hline 
         Benchmark & CM& OR & EP \\ \hline
         SLA&2.07\%& 0.77\%&1.55\%\\ \hline
         CRA&59.57\%& 3.09\% &6.38\%\\ \hline
         SNA&-7.88\%& 9.70\%&6.15\%\\ \hline
         PPJ&2.96\%& 0.24\%&7.47\%\\ \hline
    \end{tabular}
    
		}%
	\end{minipage}
\caption{\small System speed up of individual optimization over the baseline implementation in RDD.  }
    	\vspace{-3mm}
    \label{tab:perf_speedup} 
\end{table}

\begin{itemize}
    \item {\bf SLA.} There are two performance problems: CM and EP, that are detected by SODA. The revised applications are submitted to Apache Spark and become 2.07\% and 1.55\% faster than the baseline (RDD) (see Table~\ref{tab:perf_speedup}), respectively. Figure~\ref{subFigure:perf_sla} reveals that these two optimizations are not related to shuffling data size, while GC time of CM is about 2.2 times faster than the others, since cached dataset triggers a frequent GC procedure to collect JVM garbage.
    
    
    \item {\bf CRA.} All three kinds of optimizations, CM, OR and EP, can be used on this application and their performance speeds up by 59.57\%, 3.09\%, 6.38\%, respectively, according to Table~\ref{tab:perf_speedup}. In Figure~\ref{subFigure:perf_cra}, OR and EP have a positive effect on execution time and shuffling data size, while CM has speedup over execution time and does not reduce shuffling data size. Even CM has a better performance than the other two, however, the corresponding GC time is bigger. EP can reduce shuffling data size significantly but with the minimum time consumed. 
    

    \item {\bf SNA.} Table~\ref{tab:perf_speedup} shows that after applying the three optimizations, this application speeds up by -7.88\%, 9.70\%, 6.15\%, respectively. We believe two reasons are causing this worse performance (-7.88\%) of the revised application-optimized by CM: 1) this benchmark is a memory-intensive application; and 2) most of the storage memory in an executor is occupied by cached dataset, which leads to high pressure on garbage collection threads. Since SODA only handles cache memory capacity constraints and does not consider the mutual effect between storage and execution memory, such a case is difficult to be avoided. Additionally, OR has reduced shuffling data size significantly. 
    
    
    \item {\bf PPJ.} According to Table~\ref{tab:perf_speedup}, EP and CM can speed up the application by 7.47\% and 2.96\%, respectively. In Figure~\ref{subFigure:perf_ppl}, the shuffling data size has been reduced by EP from 948.8 MB to 392.2 MB while the GC time is decreased to 22.3 seconds. 
\end{itemize}

\subsection{System Overhead}
\label{subsec:overhead}

In this section, we conduct experiments in different granularity of monitoring, \eg\; monitoring no operation, partial operations suggested by SODA, or all operations involved in applications, to compare system overhead. Table~\ref{tab:sys_overhead} shows the execution time of each application with different monitoring granularity. In the partial granularity, we get profiling guidance for SLA and PPJ based on CM's suggestions, CRA and SNA based on OR's suggestions. Monitoring on all operations takes a longer time than the other two granularities. The rational reasons behind the acceptable system overhead lie in our lightweight design of {\it online} phase: 1) Enabling and customizing Spark internal {\it event} and {\it metrics} subsystems only cast needed information with a lower system overhead; 2) Exploiting data access pattern behind semantics code and DAG-based workflow provides an instrumentation guide to probe runtime system. For instance, we only consider and instrument candidate operations contributing to future ones if they are persisted in memory. It is worth mentioning that the system overhead of an application depends on its characteristic, input data size, and system configurations.






\begin{table}[ht!]
    \centering
    \resizebox{0.7\linewidth}{!}{%
    \begin{tabular}{|c|c|c|c|c|}
        \hline 
        Benchmark & Optimization & No & Partial & All\\ \hline
        SLA &CM &78.8 &87.3 &107.6 \\ \hline
        CRA &OR &240 &275.3 &532.3 \\ \hline
        SNA &OR &138 &153.4 &197.6 \\ \hline
        PPJ &CM &153.6 &176.3 &317.5 \\ \hline 
    \end{tabular}
    }%
    \caption{\small Overall comparison about System Overheads incurred by SODA}
    \vspace{-4mm}
    \label{tab:sys_overhead}
\end{table}



\section{Related Work}
\label{sec:related_work} 

A few promising programming models and platforms have been proposed through the efforts of different disciplines to accommodate the sheer size of data, such as MapReduce~\cite{dean2008mapreduce}, Apache Hadoop~\cite{hadoop2009hadoop}, Spark~\cite{zaharia2010spark} and Flink~\cite{carbone2015apache}. There is a surge of interest in optimizing data-intensive applications using semantics-aware approaches~\cite{alexandrov2014stratosphere, armbrust2015spark, guo2012spotting, jahani2011automatic, liu2012panacea, liu2018reinforcement, rheinlander2015sofa, zhang2014smarth, wang2009atomicity, zhang2017mrapid, zhang2012optimizing}. However, there is still a research gap between static and dynamic analyses to improve the performance of data-intensive systems. To the best of our knowledge, there is no optimization plugin of Scala Compiler for Spark RDD APIs, our proposed work is the first attempt in this direction.

Microsoft's Scope compiler~\cite{chaiken2008scope, garbervetsky2017static, guo2012spotting, zhang2012optimizing} automatically optimizes a data-parallel program to eliminate unnecessary code and data. It performs early filtering and calculates small derived values to minimize the amount of data-shuffling I/O based on information derived by static code analysis. There is no dynamic information involved in its optimizations. 


Spark Catalyst~\cite{armbrust2015spark} is an extensible query optimizer that leverages advanced programming language features (\eg, Scala’s pattern matching and quasi-quotes) in a novel way within the core of the Spark SQL engine. Catalyst uses a tree architecture to represent operation nodes and conduct rule-based optimizations. Finally, cost-based optimization is performed by generating multiple plans and calculating their cost to choose an optimal one. Unfortunately, it only supports Dataset and DataFrame APIs~\cite{interlandi2016optimizing, roy2019sparkcruise}. Thus, applications developed by RDD API cannot benefit from it directly. 

As to cache management, LRU caching policy is often used~\cite{zaharia2010spark, saha2015apache}. To improve cache management, several research works, namely MemTune~\cite{xu2016memtune}, LRC~\cite{yu2017lrc} and MRD~\cite{perez2018reference}, leverage directed acyclic graph (DAG), data dependency among stages, and physical schedule unit~(\ie, job and stage level) for new measurements of a data block reference. However, MemTune approach fails to answer a question of which and when each RDD will be persisted in memory. LRC updates a new reference count for each data block according to usages within a stage, however, it does not take into account the impact of data blocks spanning across multiple stages. MRD proposes a fine-grained time-locality measurement of data block reference, called reference distance. It is based on a physical schedule unit assigned by the DAG Scheduler. Nonetheless, scheduling unit orders can not reveal the real runtime executing an order to some extent. 
Our approach in SODA is a novel stage-level global cache management policy, which emphasizes two factors that would impact system performance, especially for cache behaviors: execution order of stages and data block size.

\section{Conclusions and Future work}
\label{sec:conclusion}

In this paper, we propose a semantics-aware optimization approach to assist programmers to develop and optimize an application interactively and semi-automatically. We propose three kinds of optimization strategies: cache management, operation reordering and element pruning.  Element pruning is a static rule-based model and the other two are hybrid models using static and dynamic information. To get dynamic information about data and runtime system, the online phase is developed as a piggyback monitoring tool by integrating spark internal event component, metrics system and source code profiling tools. Extensive empirical results on several real-world benchmarks using Spark RDD APIs reveal that our approach achieves better performance on the optimized code than their original implementation.

In the future work, we will extend the optimization of operation reordering to {\it map} as well as other operations. So far SODA can only take care of filter and join reordering, and help programmers choose the right operation with acceptable performance. For example, {\it reduceByKey} can replace {\it groupByKey} to reduce shuffling data size. Another promising area is to add a growing number of performance-oriented constraints to the global cache management policy. For example, we can require that all datasets needed by an operation are persisted in memory simultaneously to gain better performance.

\section{Acknowledgement} 

This work was supported in part by NSF-1836881 and NSF-1952792.




\pagebreak
\bibliographystyle{abbrv}
\bibliography{references}

\end{document}